\documentclass[twoside]{article}

\usepackage[sc]{mathpazo}
	
\usepackage[T1]{fontenc}
	
\usepackage{lmodern}
	
\usepackage{microtype}

\usepackage{amsthm}
	
\usepackage[format=hang, font=small, labelfont={bf,up}, textfont=it]{caption}
	
\usepackage{lettrine}

\usepackage{paralist}
	
\usepackage{amssymb}
	
\usepackage{gensymb}

\usepackage{csquotes}

\usepackage{hanging}

\usepackage[hmarginratio=1:1,top=32mm,columnsep=20pt]{geometry}

\usepackage{float}
	
\usepackage{abstract}

\usepackage{fancyhdr}
	
\usepackage{indentfirst}

\usepackage[hidelinks]{hyperref}
	
\usepackage{apacite}

\usepackage{titlesec}
\titleformat{\section}[block]{\large\normalfont\bfseries\centering}{\thesection.}{1em}{} 
\titleformat{\subsection}[block]{\large\bfseries}{\thesubsection.}{1em}{} 

\newcommand{\ARauthorname}{Brosnan Yuen and Mihai Sima} 

\newcommand{\ARauthorinst}{University of Victoria} 

\newcommand{\ARauthoremail}{brosnany@uvic.ca} 

\newcommand{\ARauthorthanks}{
This research was supported by a Jamie Cassels Undergraduate Research Award, University of Victoria.
} 

\newcommand{\ARtitle}{Low Cost Radiation Hardened Software and Hardware Implementation for CubeSats} 

\newcommand{\ARyear}{2018} 

\newcommand{\ARvolume}{9} 

\newcommand{\ARnumber}{1} 

\date{}

\title{
	\vspace{-15mm}
	\fontsize{18pt}{18pt}
	\selectfont
	\textbf{
		\ARtitle
	}
}

\author{
	\large
		\ARauthorname
		\thanks{
			\ARauthorthanks
		}\\[2mm]
	\normalsize
		\ARauthorinst \\
	\normalsize
		\href{mailto:\ARauthoremail}{
			\ARauthoremail
		}
	\vspace{-5mm}
}

\usepackage{graphicx}

\usepackage{amsmath}

\begin{document}

\maketitle  

\thispagestyle{fancy}


\begin{center}
	Abstract
\end{center}

\noindent CubeSats are small satellites used for scientific experiments because they cost less than full sized satellites. Each CubeSat uses an on-board computer. The on-board computer performs sensor measurements, data processing, and CubeSat control. The challenges of designing an on-board computer are costs, radiation, thermal stresses, and vibrations. An on-board computer was designed and implemented to solve these challenges. The on-board computer used special components to mitigate radiation effects. Software was also used to provide redundancies in cases of faults. This paper may aid future spacecraft design as it improves the reliability of spacecraft, while keeping costs low.


\textit{Keywords}: CubeSat; Satellite; Radiation; Radiation Hardened; Computer Memory  \\



\lettrine[nindent=0em, lines=3]{C}{ubeSats} are just scaled down versions of full-sized satellites. CubeSats are used for small payloads such as scientific experiments in space, Earth observations, and low rate telecommunications. For example, AALTO-1 CubeSat \cite{praks2015aalto} surveys mineral deposits and farmlands using an Earth observation camera. The camera allows miners to find new mining sites. Moreover, farmers can calculate the yields of their farms using the images. GeneSat-1 CubeSat \cite{kitts2007flight} is an example of a CubeSat running biological experiments in space. GeneSat-1 studies the effects of micro-gravity on E. coli bacteria. Furthermore, BRICSat-P CubeSat \cite{hurley2016thruster} has an experimental thruster design. BRICSat-P tests an electric propulsion system for future rocket designs.

 The International Space Station launches most of the CubeSats into low Earth orbit (LEO). These CubeSats orbit around 350 km to 800 km above Earth. A $10 cm \times 10 cm \times 10 cm$ CubeSat costs around \$30,000 USD \cite{heidt2000cubesat} to build and \$100,000 USD to launch. On the other hand, a 100kg full-sized satellite costs \$100 million USD \cite{bearden2001small} to build and to launch. Costs for constructing and launching CubeSats are significantly lower than for full-sized satellites. Therefore, CubeSats are low-cost alternatives to full-sized satellites. In conclusion, these low-cost CubeSats have been adopted by many universities across the world as an exploration tool.

The CubeSat's primary objective is the payload execution. Most payloads contain scientific experiments or sensor equipment. The CubeSat uses the on-board computer (OBC) to execute the payload and to retrieve sensor data. Moreover, the OBC has a microcontroller (MCU) and a memory storage. The MCU processes data, while the memory store stores the sensor data and the program data. After processing the data, the data is sent to Earth using amateur radio. Space-grade components such as the \$950 VA10820 \cite{VA10820}, \$2900 MSP430FR5969-SP \cite{MSP430FR5969}, \$4000 5962H9853702VXC-CTEST \cite{5962H9853702VXC_CTEST}, and \$5100 HXNV01600AEN \cite{HXNV01600AEN}  are very expensive. As CubeSat projects have low budgets of \$30,000, CubeSat designers cannot afford to use space grade components. As a result, CubeSat designers use commercial off the shelf (COTS) components as COTS components are far cheaper. However, COTS components are less robust when compared to space grade components. COTS components have lower tolerance to radiation, vibrations, and thermal stresses. This lower tolerance limits the CubeSat's lifespan to two years in LEO. Radiation poses a large risk to the OBC as memory regions of the OBC could get corrupted. Moreover, radiation could cause permanent damage to the OBC's integrated circuits (ICs). Permanent damage results in an unrecoverable failure.

This paper is organized into a literature review, design overview and requirements, implemented designs, and conclusion. The literature review gives detailed background information about the space environment and it compares this paper to other research papers. Design overview and requirements gives an outline of the implemented designs. The paragraph above states the problems of the COTS components. Solutions of these problems are found in the implemented design sections of this paper. The design implementations use special COTS components in conjunction with robust software designs. These special COTS components have a high tolerance to radiation, temperature, vibrations, and noise. Furthermore, these special COTS components meet the design requirements while keeping the costs low. In addition to the hardware solutions, software designs also mitigate the problems caused by radiation. In conclusion, this paper will assist CubeSat designers in advancing space exploration and scientific experiments.

\section*{Literature Review}

This literature review shows the relevant background information about the space environment. Total Ionizing Dose (TID), Single Event Upsets (SEUs), Single Event Latch-ups (SELs), and Single Event Gate Ruptures (SEGRs) measure the impact of radiation on ICs. Firstly, TID measures the total amount of dose an IC could receive before an unrecoverable failure. TID quantifies the lifetime of an IC in space. Secondly, SEUs measure the number of corrupted bits and computations caused by radiation. However, SEUs do not have any long-term  effects as cold reboots resolve SEUs. On the other hand, SELs could have long-term  effects. SELs cause high current states, which may cause permanent burnouts in IC. Lastly, SEGRs cause parts of an IC to rupture. SEGRs guarantee permanent loss of functionality in an IC.

Other CubeSat designers have tackled the radiation problem using hardware solutions. Shields-1 CubeSat \cite{thomsen2015shields} uses special shielding to protect the COTS components against radiation. Shields-1 uses Z-grade Al/Ta which has 30\% higher shielding effectiveness over standard Al. \citeauthor{austin2017cubesat}'s article \citeyear{austin2017cubesat} shows radiation mitigation techniques for COTS components. Their article recommends measuring TIDs and SEUs of COTS components. This allows CubeSat designers to choose the best COTS components. \citeauthor{austin2017cubesat} also recommend the usage of ferromagnetic random access memory (FRAM) and watchdog timers. These components increase the reliability of the CubeSats. Unlike \citeauthor{austin2017cubesat}, this paper uses software solutions in addition to hardware solutions. The software solutions in this paper increase the reliability of the CubeSats as software provides a cost-effective way of mitigating radiation effects.

\section*{Design Overview and Requirements}
\begin{figure}[h!]
\centering
\includegraphics[width=0.55\textwidth]{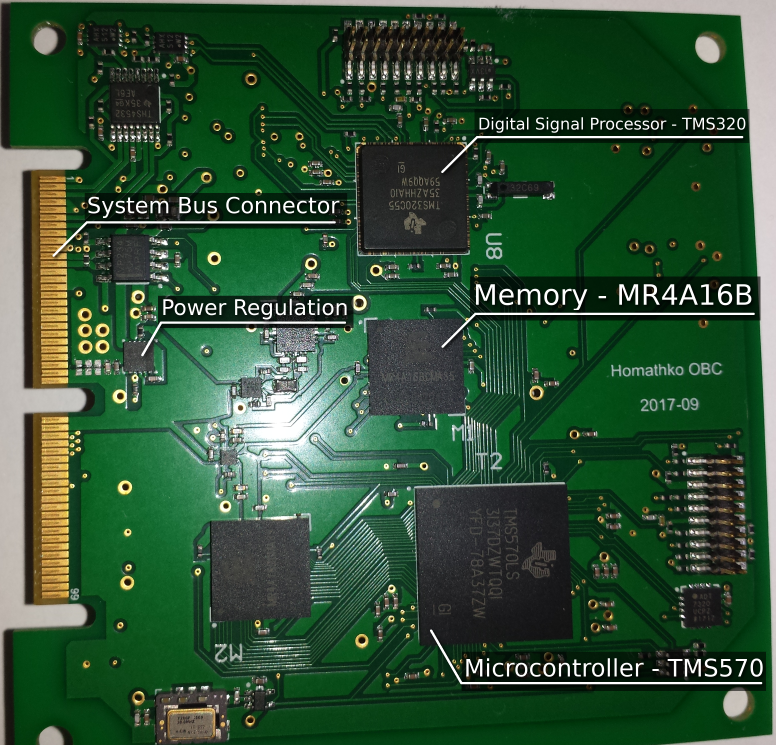}
\caption{\label{fig:pcb} The implemented OBC's printed circuit board (PCB).}
\end{figure}

The OBC is the brains of the CubeSat because the OBC controls all of the CubeSat's functions. Figure \ref{fig:pcb} shows the implemented OBC's printed circuit board (PCB). The OBC's PCB has a budget requirement of \$2000 USD. The MCU is the TMS570LC4357. The memory storage is the MR4A16BCMA35. TMS570 uses MR4A16BCMA35 for external memory storage. Moreover, the system bus provides power and signal lines to the OBC. As a result of the system bus, TMS570 is able to control other systems. For example, Attitude Determination and Control System (ADCS) has GPS, magnetometers, gyroscopes, sun sensors, and Earth horizon sensors. ADCS determines the position and orientation of the CubeSat using the sensors. ADCS also adjusts the orientation of the CubeSat using magnetic solenoids and reaction wheels. Furthermore, OBC has a TMS320C5535 digital signal processor. TMS320C5535 processes the low rate telecommunications on the CubeSat. Telecommunications allow the ground station to communicate with the TMS570. Two TPL5010-Q1 external watchdog timers are placed on the OBC. Watchdog timers reset the OBC in an event of a failure. These low-cost COTS components are the solutions to the design requirements problem. Unlike other COTS components, these components have high radiation tolerances and they operate across large temperature ranges. Moreover, they are space flight proven. In conclusion, these COTS components cost less while ensuring the CubeSat survives the space environment.

Aside from the hardware descriptions, there are functional requirements. Telemetry, payload control, power control, data handling, timekeeping, and memory integrity are examples of the OBC's functional requirements. Telemetry is the act of collecting remote sensor data. As a payload is used on all CubeSat missions, payload control ensures the payload is executing properly. Power control ensures brown-outs do not happen. Data handling involves sensor processing, telecommunications, and system monitoring. Memory integrity protects the memory from data corruption.

\begin{table}[h!]
\begin{center}
    \begin{tabular}{ | l | l |}
    \hline
    Type of radiation & Dose rate ($ \frac{MeV cm^{2}}{g}$)  \\ \hline

Heavy ion & $4.75 \times 10^{2}$ \\ \hline

Proton & $9.62 \times 10^{2}$  \\ \hline

Electron & $1.359 \times 10^{3}$ \\ \hline

Total & $2.796 \times 10^{3}$   \\ \hline

    \end{tabular}
\end{center}
  \caption{Dose rates for solar maximum at $8 \times 10^{5}$ m LEO \protect\cite{hussmann2016reliable}.} \label{tab:dose_rates}
\end{table}

 Hardware requirements were made for LEO as they ensure the OBC's survivability. Table \ref{tab:dose_rates} shows the dose rates of a CubeSat at $8 \times 10^{5}$ m LEO with $2 \times 10^{-3}$ m Al shielding. Table \ref{tab:dose_rates}  assumes a mission duration of $6.307 \times 10^{7}$ s (2 years). At the end of the mission, the total received dose of a $1 cm \times 1 cm$ component is $2.825 \times 10^{3}$ rad (Si) as shown in Eqn. \eqref{eq:TID}. Therefore, components must survive TIDs of $2.825 \times 10^{3}$  rad (Si). Moreover, components are susceptible to SEUs. Components must also withstand a few SEUs per hour. DIETR \cite{dietr} specifies a thermal-vacuum requirement. OBC must withstand a vacuum of $5\times10^{-4}$ Torr. OBC must also have an out-gassing requirement \cite{outgassing}. This confines out-gassing to 1\% of the total mass. In addition, OBC must withstand thermal cycles from $-20 ^{\circ}$C to $70^{\circ}$C.  DIETR also specifies a Launch Environment Tests requirement \cite{launch}. OBC must withstand a quasi-static acceleration test of 12g. The OBC must also withstand random vibrations of $1 \times 10^{-1} \frac{g^{2}}{Hz}$ at $2 \times 10^{2} $ Hz. OBC's PCB is confined to a 10 cm by 10 cm size. 

For the software requirements, a real time operating system (RTOS) is needed. A RTOS schedules tasks with high timing accuracy as the RTOS allows microsecond control of the CubeSat. Moreover, RTOS also manages memory allocations and hardware drivers. A file system is also needed to protect the data from radiation. The file system must have multiple copies of the file system structure. The file system must use an error correction code (ECC) for data recovery. ECC protects data by encoding the data. Encoded data contains multiple copies of the original data. The numerous data copies provide redundancies in case of failures. The file system must be scrubbed within a time interval in order to recover the corrupted data.

\section*{Implemented Memory Design}

The OBC requires a reliable place to store sensor data and program data. MR4A16BCMA35 was chosen as the nonvolatile memory storage for the MCU. Each MR4A16BCMA35 stores 2MB of data. MR4A16BCMA35 is low-cost as MR4A16BCMA35 costs around \$40 USD \cite{DIGIKEY_MRAM}. MR4A16BCMA35 is cheap when compared to other memory storages. For example, 5962H9853702VXC-CTEST costs \$4000 \cite{5962H9853702VXC_CTEST} and \\
HXNV01600AEN costs \$5100 \cite{HXNV01600AEN}. MR4A16BCMA35 \cite{MR4A16B} uses magneto-resistive random access memory (MRAM) technology. In MRAM, bits are stored in magnetic fields of the magnetic tunnel junctions. A magnetic tunnel junction has two ferromagnets and one insulator. The insulator separates the two ferromagnets. One ferromagnet is permanent, while the other ferro-magnet is a free layer. Orientation of the magnetic field in the free layer determines the bit's value. In order to write a bit, a large current is required to change the orientation of the free layer.

Radiation in LEO consists mostly of heavy ion bombardments. As NAND memory uses charge pumps to store bits, NAND memory is susceptible to heavy ions bombardments. Heavy ions bombardments corrupt bits in NAND memory by changing values in the charge pumps. On the other hand, MRAM is resistant to heavy ions bombardments as MRAM uses magnetic fields to store bits. Heavy ions bombardments cannot generate large surges of current. Therefore, heavy ions bombardments cannot corrupt bits in MRAM by changing the magnetic fields. As a result, MRAM has large SEU and SEL damage thresholds. In conclusion, MRAM is resistant to radiation as MRAM uses magnetic fields to store bits. Table \ref{tab:dose_rates} shows dose rates of components in LEO. The required TID in LEO is calculated below using the dose rates in Table \ref{tab:dose_rates}. Then MRAM's TID is compared to the required TID in order to prove MRAM's survivability.

Let $ LEO_{doserateperarea} = $ total dose rate per area of an object in LEO as shown in Table \ref{tab:dose_rates}

Let $ MRAM_{area} = $ the area of MRAM 

Let $ LEO_{doserate} = $ the dose rate of an object in  $8 \times 10^{5}$ m LEO
$$ MRAM_{area} = 1 cm^{2}$$
$$ LEO_{doserateperarea} = 2.796 \times 10^{3} \frac{MeV \cdot cm^{2}}{g \cdot  s}$$
$$ 2.796 \times 10^{3} \frac{MeV \cdot cm^{2}}{g \cdot  s} \times \frac{1}{1 cm^{2}}  =  2.796 \times 10^{3} \frac{MeV}{g \cdot  s}$$
$$  2.796 \times 10^{3}  \frac{MeV}{g \cdot  s} \times \frac{1000 g}{1 kg} \times \frac{1.602 \times 10^{-13} J}{ 1 MeV} = 4.48 \times 10^{-7} \frac{J }{ kg \cdot s}$$
$$ LEO_{doserate} = 4.48 \times 10^{-7} \frac{J }{ kg \cdot s} \times  \frac{100 rad \cdot kg}{ J  }  $$
$$ LEO_{doserate} = 4.48 \times 10^{-5}  \frac{ rad  }{ s  }  $$

Let $T = $ mission duration in seconds. $6.307 \times 10^{7}$s (2 years).

Let $LEO_{TID} = $ required TID to survive in LEO for $6.307 \times 10^{7}$s (2 years)

Let $MRAM_{TID} = $ MRAM's TID as shown in Table \ref{tab:mram_prop}
$$ T = 6.307 \times 10^{7} s $$
\begin{equation} 
LEO_{TID} =  LEO_{doserate} \times T 
\label{eq:TID} 
\end{equation} 
$$ LEO_{TID} =  4.48 \times 10^{-5}  \frac{ rad  }{ s  } \times  6.307 \times 10^{7} s $$
$$ LEO_{TID} = 2.825 \times 10^{3} rad (Si)$$
$$ MRAM_{TID} = 4 \times 10^{4} rad (Si) $$
$$ MRAM_{TID} >> LEO_{TID} $$
\begin{table}[!h]
\centering
\begin{tabular}{| l | l |}
\hline
Parameter      & Limits                                       \\ \hline
Data retention & 20 Years                                     \\ \hline
TID            & $4 \times 10^{4}$ rad (Si)    \\ \hline
SEU            & \textgreater $1 \times 10^{2} \frac{MeV \cdot cm^{2}}{mg} $  \\ \hline
SEL            & \textgreater $8.4 \times 10^{1} \frac{MeV \cdot cm^{2}}{mg} $  \\ \hline
Access time    & $3.5 \times 10^{-8}$ s                                       \\ \hline  
ECC            & 7 bits of parity per 64 bits          \\ \hline    

H field tolerance & $ 8 \times 10^{3} \frac{A}{m} $ \\ \hline 

\end{tabular}
\caption{Properties of Everspin MRAM \protect\cite{MR4A16B} \protect\cite{heidecker2013mram} \protect\cite{zhang2016total}.} \label{tab:mram_prop}
\end{table}

Table \ref{tab:mram_prop} shows the properties of Everspin MRAM. MRAM lasts for 20 years within a temperature range of $-40 ^{\circ}$C to $85^{\circ}$C \cite{MR4A16B}. Furthermore, MRAM has a TID of $4 \times 10^{4}$ rad (Si) \cite{zhang2016total}. MRAM's TID of $4 \times 10^{4}$ rad (Si)  is far greater than the required TID of $2.825 \times 10^{3}$ rad (Si) calculated in Eqn. \eqref{eq:TID}. As a result of a large TID tolerance, MRAM survives the radiation in LEO. In spite of MRAM's TID being overkill, MRAM is the only component that has a known TID and costs less than \$40 USD. Other components such as FRAM, NAND flash, and NOR flash have unknown TIDs. On the other hand, 5962H9853702VXC-CTEST and HXNV01600AEN have TIDs of $1 \times 10^{6}$ rad (Si) but cost \$4000 and \$5100 respectively. MRAM has virtually unlimited read/write endurance as MRAM does not wear out. MRAM's access time of $3.5 \times 10^{-8}$ s matches MCU's SDRAM's access time. MRAM's SEU is greater than $> 1 \times 10^{2} \frac{MeV \cdot cm^{2}}{mg}$ \cite{heidecker2013mram}. Therefore, MRAM tolerates SEUs. MRAM mitigates bit flips caused by SEUs as MRAM has high SEU tolerance.

SEUs determines the bit error rate (BER) of memory. BER is important for MRAM as BER is used to determine the total number of bits errors over a time period. Using BER from \citeauthor{heidecker2013mram}'s \citeyear{heidecker2013mram} study, the total number of bit errors was calculated below for 4MB MRAM over $7.3 \times 10^{2}$ days (2 years). Total number of bit errors shows the MRAM's data integrity. Note: An upset is a bit error.

Let $p_{upset} = $ probability of an upset per bit day 
$$ p_{upset} = 1 \times 10^{-10} \frac{upsets}{bit \cdot day}  $$
Let $p = $ probability of an upset for 1 bit in $7.3 \times 10^{2}$  days
$$ p = 1 \times 10^{-10} \frac{upsets}{bit \cdot day} \times 7.3 \times 10^{2} days \times 1 bit  $$
$$ p = 7.3 \times 10^{-8} upsets  $$
Calculations below show a PDF of bit errors for a 4MB MRAM with parity bits.

Let $N = $ number of bits in memory

Let $k = $ number of corrupted bits

Let $f = $ binomial distribution
$$ N = 3.55 \times 10^{7} bits $$
\begin{equation}
f = \binom{N}{k} p^{k} (1-p)^{N-k}
\label{eq:binominal}
\end{equation}
\begin{figure}[h!]
\centering
\includegraphics[width=0.7\textwidth]{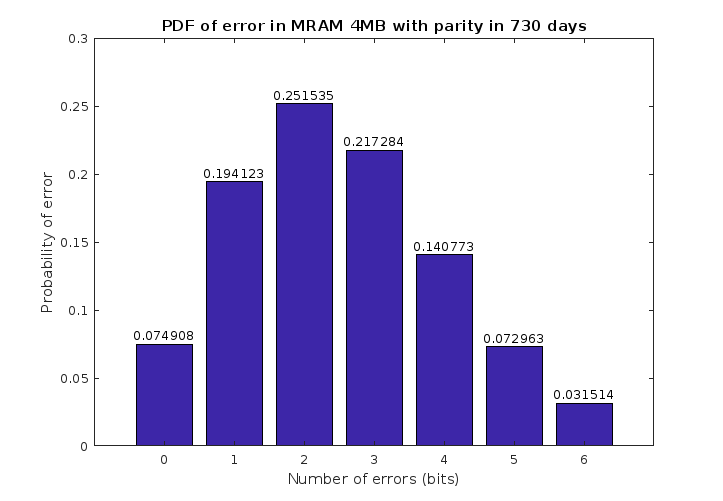}
\caption{\label{fig:mram_pdf1}PDF of error in MRAM's 4MB memory in 730 days.}
\end{figure}

\newpage

Figure \ref{fig:mram_pdf1} shows the binomial distribution of bit errors generated from Eqn. \eqref{eq:binominal}. Mean bit errors are 2 bits over $7.3 \times 10^{2}$ days. As a result of low bit errors, MRAM fully corrects the bit errors. This results in zero bit errors when MRAM is turned on.

However, there is a case where MRAM is turned off. During rocket launch, MRAM is powered off for 1 day. When MRAM is turned off, MRAM cannot correct bit errors until MRAM is turned back on. MRAM's ECC uses 7 bits of parity for every 64 bits \cite{MR4A16B}. MRAM's ECC recovers a 1 bit error in a 71 bit block. MRAM's ECC protects against single bit flips caused by radiation. The calculations below show the probability of failure when MRAM is turned off for 1 day. The probability of failure determines the data's integrity during rocket launch.

Let $p = $ probability of an upset for 1 bit in 1 day
$$ p = 1 \times 10^{-10} \frac{upsets}{bit \cdot day} \times 1 day \times 1 bit  $$
$$ p = 1 \times 10^{-10} upsets $$
Let $N = $ number of bits per block
$$ N = 71 bits $$
Used Eqn. \eqref{eq:binominal} to solve for the probability $P_{71bits}$ when $ k > 1 $. 

Let $P_{71bits} = $ probability of ECC not correcting bit errors in 71 bits in 1 day
$$ P_{71bits} = f(k > 1) $$
$$ P_{71bits} = 6.66 \times 10^{-16} $$
There are $ 5 \times 10^{5} $ blocks of 71 bits for the total memory capacity (4 MB).
$$ N = 5 \times 10^{5} blocks$$
Used Eqn. \eqref{eq:binominal} to solve for the probability $P_{4MB}$ when $ k > 0 $.

Let $P_{4MB} = $ probability of ECC not correcting bit errors in 4MB in 1 day 
$$ P_{4MB} = f(k > 0) $$
$$ P_{4MB} = 3.33 \times 10^{-10} $$

The probability of failure when MRAM is turned off for 1 day is $ P_{4MB} = 3.33 \times 10^{-10} $. As a consequence of a small probability of failure, MRAM's ECC repairs the bit errors caused by radiation. In conclusion, MRAM's ECC preserves data integrity during rocket launch.

\section*{Implemented Microcontroller Design}

A MCU is required to control the CubeSat. TMS570 is a MCU series \cite{TMS570LC4357}. TMS570 is low-cost as TMS570 costs around \$60 USD \cite{DIGIKEY_TMS570}. TMS570 is inexpensive when compared to other MCUs such as \$950 VA10820 \cite{VA10820} and \$2900 MSP430FR5969-SP \cite{MSP430FR5969}. TMS570 uses dual lockstep CPUs. The dual lockstep CPUs are identical. Each CPU on the TMS570 independently executes the exact same instruction. If the exact same instruction produces different results, then both CPUs will re-execute the instruction again. Dual lockstep CPUs protect against SEUs as they prevent incorrect computations. TMS570 also has eFuses with parity bits. In eFuses, hardwired fuses store data. eFuses protect against radiation as radiation cannot damage fuses. TMS570 also has self testing capabilities to detect errors in hardware. At boot-up, the TMS570 checks each built-in module. Self testing enables the TMS570 to troubleshoot errors inflight.

TMS570 operates within a temperature range of $-40 ^{\circ}$C to $125^{\circ}$C, thus satisfying the temperature requirements. TRIUMF is a particle accelerator used for radiation testing. Radiation testing at TRIUMF determined the TMS570's TID of $5 \times 10^{3}$ rad (Si). Therefore, TMS570 survives the required TID of $2.825 \times 10^{3}$  rad (Si). On the other hand, VA10820 and MSP430FR5969-SP have TIDs of $3 \times 10^{5}$ rad (Si) and $5 \times 10^{4}$ rad (Si) respectively. VA10820 and MSP430FR5969-SP have larger TIDs, but they cost more than the TMS570. The TMS570's 4MB internal memory is used for program data and the boot loader. Moreover, TMS570's internal memory uses 64/72 bit hamming for ECC. TMS570's ECC corrects 1 bit for every 72 bits. During rocket launch, TMS570 is powered off for 1 day. When TMS570 is turned off, TMS570 cannot correct bit errors until TMS570 is turned back on. The calculations below are for the probability of bit errors when TMS570 is turned off for 1 day. The bit errors determine TMS570's data integrity during rocket launch.

Let $H = $ number of hours

Let $p = $ the probability of error in a bit 

Let $N = $ number of bits 

Let $k = $ number of corrupted bits 

Let $f = $ binomial distribution 
\begin{equation}
p = 1 - e^{-H  (9.0813 \times 10^{-6})}
\label{eq:tms570_bit_error}
\end{equation}
$$ N = 72 bits$$
$$ H = 24 hours$$
\begin{figure}[h!]
\centering
\includegraphics[width=0.7\textwidth]{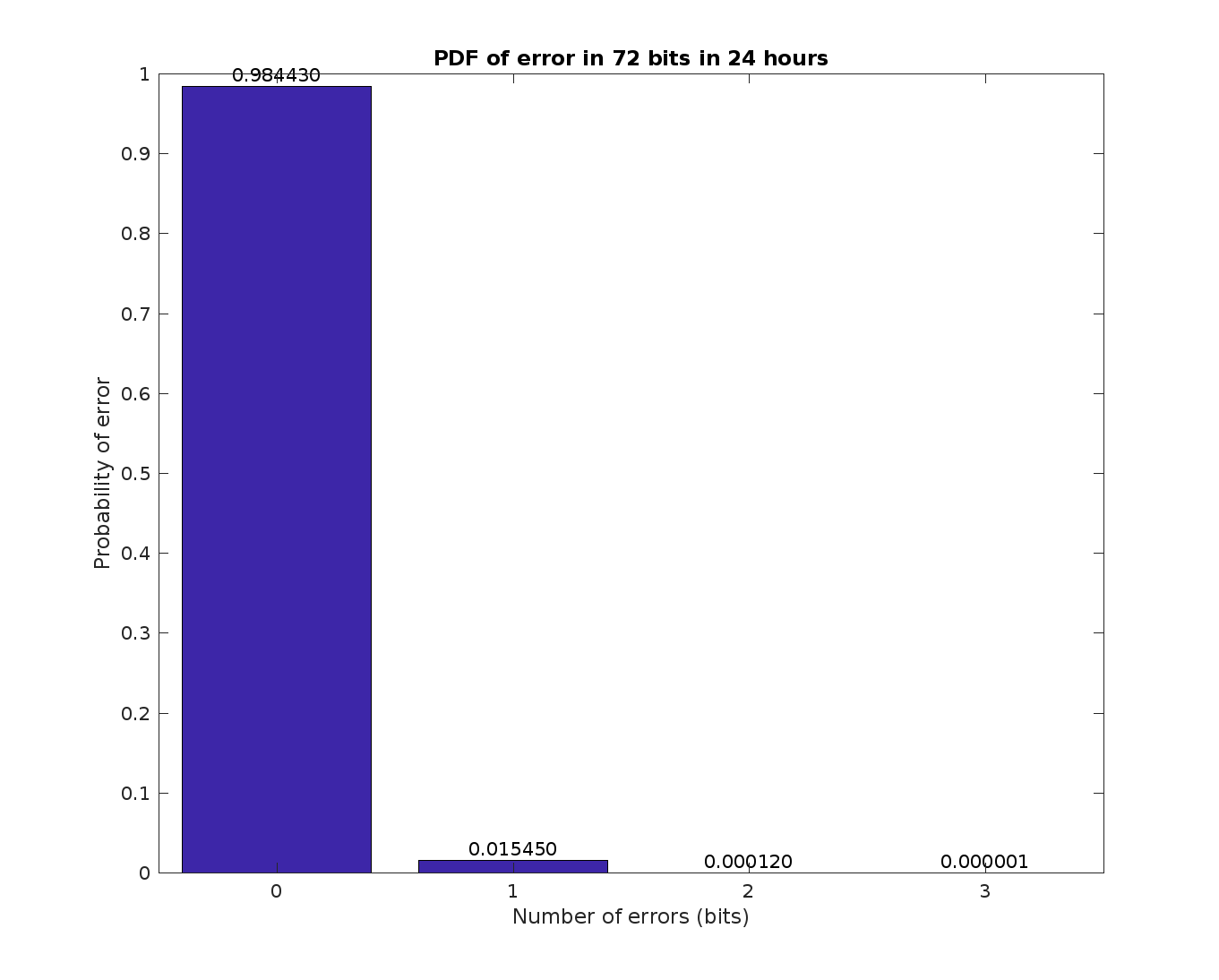}
\caption{\label{fig:tms570_72}PDF of error in 72 bits in 24 hours.}
\end{figure}
Let $P_{72bits} = $ probability of ECC not correcting bit errors larger than 1 bit error for 72 bits
$$ N = 79  blocks$$
$$ P_{72bits} = f(k > 1)$$
$$ P_{72bits} =   1.2 \times 10^{-4} $$

Figure~\ref{fig:tms570_72} shows the probability of ECC not correcting bit errors larger than 1 bit error for 72 bits. Bit error calculations used Eqn. \eqref{eq:binominal} and Eqn. \eqref{eq:tms570_bit_error} \cite{hussmann2016reliable}. The bit error probability is very low as $ P_{72bits} = 1.2 \times 10^{-4} $.

The boot-loader is a critical piece of software. If radiation corrupts the boot-loader then the entire CubeSat is useless. Therefore, the probability of error for the boot-loader is calculated below using Eqn. \eqref{eq:binominal}. The boot loader uses 5000 bits or 79 blocks of TMS570's internal memory. Each block has 72 bits. The probability of error determines boot-loader's survivability during rocket launch.

Let $ P_{5000bits} = $ probability of error in 5000 bits
$$ P_{5000bits} =  f(k > 0)$$
$$ P_{5000bits} = 9.4 \times 10^{-3} $$
The probability of error in the boot loader during launch is $ P_{5000bits} = 9.4 \times 10^{-3}$. As the probability of error is small, the boot-loader is safe from corruption during rocket launch.

Aside from the boot-loader, TMS570 also stores program data. Program data is susceptible to radiation during rocket launch. Therefore, bit errors in the program data will be analyzed below. Bit errors for TMS570's 4MB memory are calculated below using Eqn. \eqref{eq:binominal} and Eqn. \eqref{eq:tms570_bit_error}. The bit errors determine percentage of corrupted regions in the program data.

$$ H = 24 hours$$

$$ p = 7.3 \times 10^{-8} $$

$$ N = 3.6 \times 10^{7} bits $$

\begin{figure}[h!]
\centering
\includegraphics[width=0.7\textwidth]{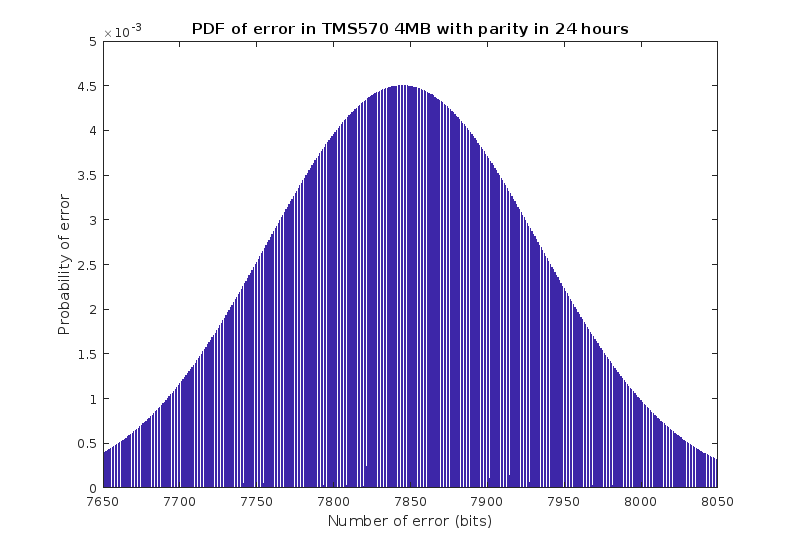}
\caption{\label{fig:tmspdf}PDF of TMS570's 4MB memory with parity in 24 hours.}
\end{figure}

 Figure~\ref{fig:tmspdf} shows the bit errors for the TM570's 4MB memory during a 24 hour period. Standard deviation is 55 bit errors and the mean is 7860 bit errors. Due to large bit errors in the TM570's 4MB memory, a file system is needed to mitigate the bit errors. The file system design is found in the software design section.

 \section*{Implemented Software Design}

In addition to the hardware, software also mitigates radiation. TMS570 uses FreeRTOS, a lightweight and modular RTOS. FreeRTOS handles task scheduling, interrupt scheduling, memory allocation, and hardware driver management. For task and interrupt scheduling, FreeRTOS allows preemptions and static memory allocations, which increase the timing accuracy of tasks and interrupts. FreeRTOS allows the OBC to react faster to faults and errors. Furthermore, FreeRTOS uses a round robin scheduler to control the execution of tasks. For each task with the same priority, the tasks receive the same CPU execution times. Along with tasks, interrupts are used to warn the MCU of high priority tasks. For example, if the battery system is running out of energy then an interrupt fires. After the interrupt fires, the MCU focuses solely on the interrupt and turns off certain systems to conserve energy. As FreeRTOS manages memory allocation, FreeRTOS runs directly on the MRAM instead of the TMS570's internal memory. Running on the MRAM reduces bit errors caused by radiation.

\begin{figure}[h!]
\centering
\includegraphics[width=0.7\textwidth]{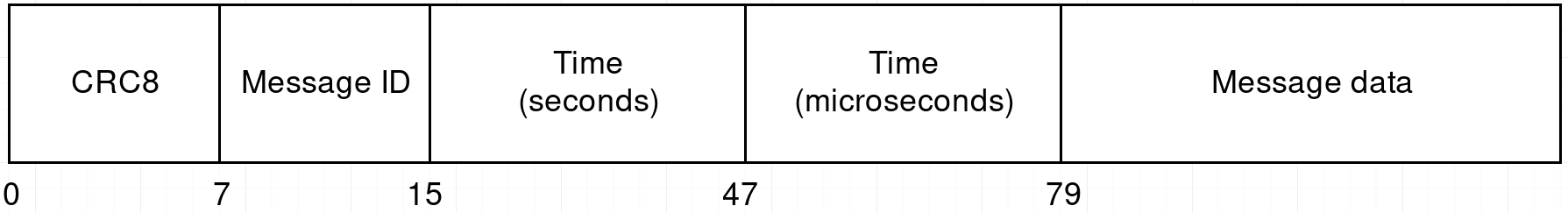}
\caption{\label{fig:system_bus}Bit arrangement in the modified queue system.}
\end{figure}

Queues in FreeRTOS are used for interprocess communications between different tasks. For example, one task sends sensor data to a buffer queue. Another task receives sensor data from the buffer queue and processes the sensor data. Moreover, queues have built-in mutexes to eliminate race conditions between tasks. However, the default queues in FreeRTOS are not very robust. Therefore, the default queues will be modified. Figure~\ref{fig:system_bus} shows the features of the modified queue system. Each queue element stores an 8 bit cyclic redundancy check (CRC8). CRC8 verifies the element's data integrity as CRC8 detects up to 8 bit errors in a element. If CRC8 detects a corrupted element, then the element is discarded. The 8 bit message ID represents the sensor's label. For example, sensor 1 will have a message ID of 1. Sensor 5 will have a message ID of 5. Message ID enables the receiving task to distinguish between sensors. Messages also have time stamps associated with them. Time stamps take up to 64 bits and are measured in seconds and microseconds. A time stamp determines the sample time of the data.

File systems protect against data corruption as radiation corrupts data. The file system uses Bose Chaudhuri Hocquenghem (BCH) codes \cite{chien1964cyclic}, a type of ECC. BCH codes protect the data by encoding the data. In the file system, data is placed into blocks. BCH codes encode each data block. Each encoded block contains parity bits. Parity bits are copies of the original data. When an encoded block gets corrupted, BCH codes recover the original data using the parity bits in the encoded block. As a consequence, BCH codes repair the errors caused by radiation. Thus, BCH codes protect against radiation. Moreover, BCH codes allow designers to determine the number of bits the BCH codes recovers. For every $mt$ parity bits in a BCH code, the BCH code recovers up to $t$ bits. 

The BCH code requires calculations to determine the optimal number of $mt$ parity bits. If $mt$ is too small, then the file system cannot recover from bit errors. If $mt$ is too large, then the parity bits take up too much space. Calculations below determine the optimal number of $mt$ parity bits for the TMS570's 4MB memory. Calculations assume the file system has 4095 bit blocks with a scrubbing period of 6 hours.

Let $N = $ number of bits in a block 

Let $H = $ scrubbing period in hours

Let $m = $ minimal polynomial over the field $GF(q^{m})$

Let $p = $ probability of a bit flip 

Let $f = $ binomial distribution

Let $t = $ number of recoverable bit errors in a block

Let $R = $ number of parity bits in the block 
$$ m = 12 $$
$$ N = 2^{m} - 1 $$
$$ N = 2^{12} - 1 = 4095$$
$$ H = 6 hours$$

Using Eqn. \eqref{eq:tms570_bit_error}. The probability of one bit flip in 6 hours is calculated.
$$ p = 5.448 \times 10^{-5} $$
Using Eqn. \eqref{eq:binominal} to create a function for every $t$ input there is a $P_{4095bits}$ output.

Let $P_{4095bits} = $ probability of error in a 4095 bit block
$$ P_{4095bits} =  f(k > t) $$
Let $N = $ number of 4095 bit blocks in TMS570's 4MB memory
$$ N = 7815 $$
Using Eqn. \eqref{eq:binominal} again for the total probability of error in TMS570's 4MB memory.

Let $P_{error} = $ total probability of error in the entire TMS570's 4MB memory
$$P_{error} = f(k > 0)$$

\begin{figure}[h!]
\centering
\includegraphics[width=0.7\textwidth]{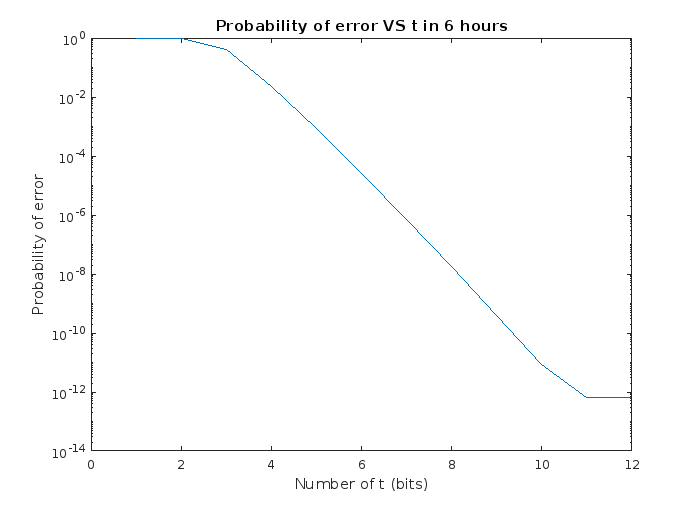}
\caption{\label{fig:TMS570_FS}Probability of error vs number of t bits in TMS570's 4MB memory.}
\end{figure}

Figure \ref{fig:TMS570_FS} shows the probability of error in TMS570's 4MB memory vs number of $t$ bits. For the probability of error $P_{error} < 10^{-10}$, select $ t = 9$ bits. The $m = 12$ is determined by the minimal polynomial over the field $GF(q^{m})$.
$$ R = mt $$
$$ R = 12 \times 9 bits = 108 bits$$
Therefore, the file system requires $R=108$ parity bits for  $P_{error} < 10^{-10}$. BCH code recovers up to $t=9$ bit errors for every 4095 bit block in the TMS570's 4MB memory.

The optimal number of parity bits for MRAM's 4MB memory was calculated below. Furthermore, the scrubbing period was changed to $H = 24 hours$. The probability of one bit flip for MRAM is changed to $p = 10^{-10}$. Other parameters remained the same.
$$ H = 24 hours $$
$$ p = 10^{-10} $$
\begin{figure}[h!]
\centering
\includegraphics[width=0.7\textwidth]{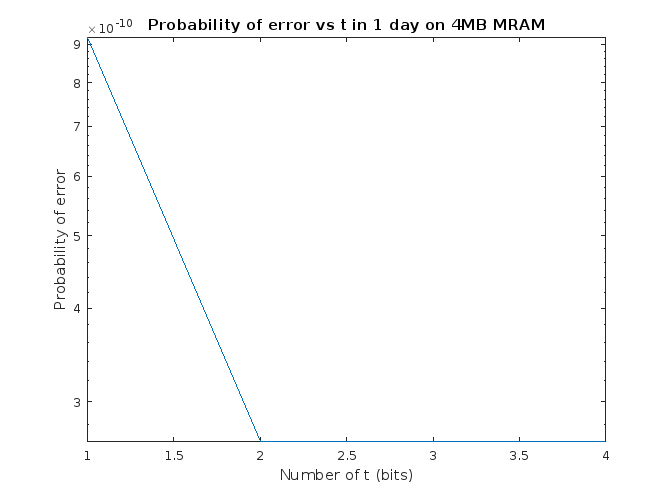}
\caption{\label{fig:MRAM_FS}Probability of error vs number of t bits in MRAM's 4MB memory.}
\end{figure}

Figure \ref{fig:MRAM_FS} shows probability of error in MRAM's 4MB memory vs number of t bits. For an error probability of $P_{error} < 10^{-10}$, select $ t = 2$ bits. 
$$ R = 12 \times 2 bits = 24 bits$$
Therefore, the file system requires $R=24$ parity bits for $P_{error} < 10^{-10}$. BCH code recovers up to $t=2$ bit errors for every 4095 bit block in MRAM's 4MB memory.

\section*{Implemented Watchdog Design}

\begin{figure}[h!]
\centering
\includegraphics[width=0.4\textwidth]{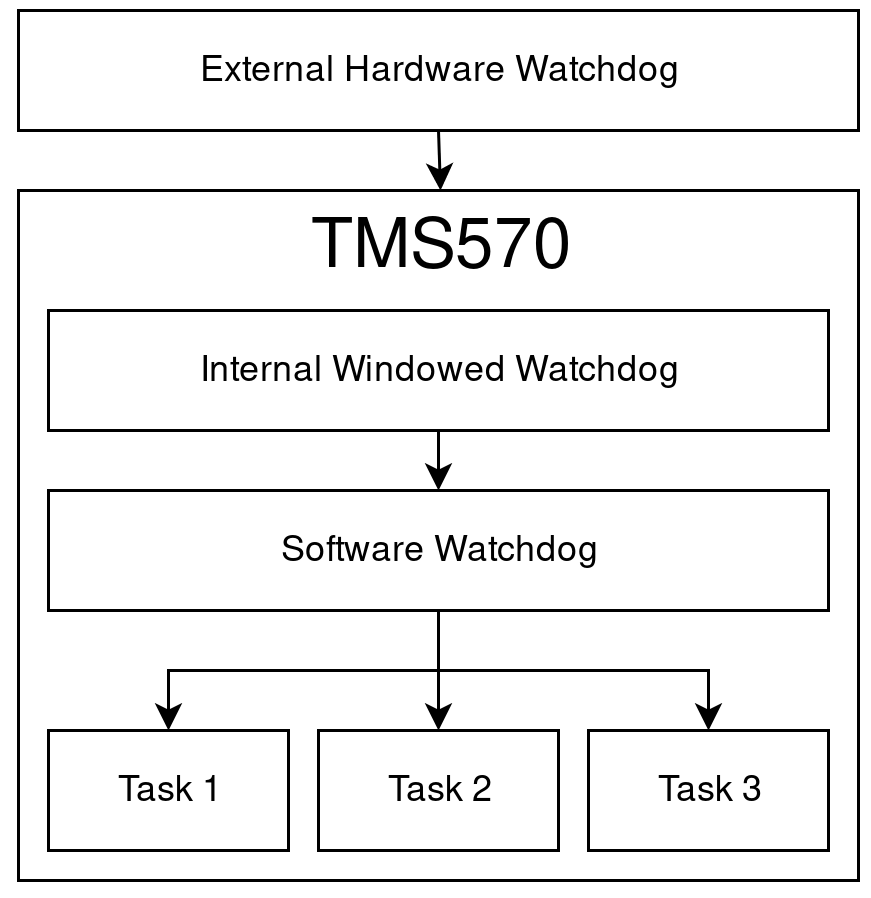}
\caption{\label{fig:watchdogs_layers}Layers of watchdogs on the OBC.}
\end{figure}

The space environment causes many faults in ICs. However, watchdogs mitigate the faults by restarting ICs. Thus, watchdogs increase the reliability of the OBC. TMS570 has, TPL5010-Q1, a dedicated external hardware watchdog is shown in Figure~\ref{fig:watchdogs_layers}. At \$2 USD \cite{DIGIKEY_TPL5010}, TPL5010-Q1 is low-cost. TPL5010-Q1 functions within a temperature range of $-40 ^{\circ}$C to $125^{\circ}$C and is vibration resistant. TMS570 periodically sends done signals to TPL5010-Q1. TPL5010-Q1 will force a power reset on the TMS570 if TPL5010-Q1 does not receive a done signal. TPL5010-Q1 ensures the TMS570 recovers from faults. Moreover, TMS570 has a built-in windowed watchdog. The windowed watchdog monitors FreeRTOS for a done signal and will force a cold reboot on the TMS570 if it does not receive a done signal. The windowed watchdog adds redundancy to the watchdog setup. Furthermore, a software watchdog monitors the individual tasks in FreeRTOS. The software watchdog activates periodically. When the software watchdog activates, the software watchdog checks the tasks' time stamps. Each task has a time stamp, which contains the last time the task fed the software watchdog. If a task's time stamp is not within the time interval, then the task will be restarted. Restarting a task clears the previous state and memory of the task. Thus, the software watchdog increases reliability of each individual task.

 \citeauthor{beningo2010review} \citeyear{beningo2010review} shows the probabilities of watchdogs successfully recovering from faults. Calculations below show the probabilities of success for the entire watchdog system. The probabilities of success for recovering from a fault quantify the reliability of the OBC.

Let $P(EWDG) = $ probability of external hardware watchdog successfully recovering from a fault

Let $P(WWDG) = $ probability of windowed watchdog successfully recovering from a fault 

Let $P(SWDG) = $ probability of software watchdog successfully recovering from a fault 
$$ P(EWDG) = 0.85 $$
$$ P(WWDG) = 0.85 $$
$$ P(SWDG) = 0.7 $$ 
Let $P(EWDG \cup WWDG) = $ probability of external hardware watchdog or windowed watchdog successfully recovering from a fault
$$ P(EWDG \cup WWDG) =  P(EWDG) + P(WWDG)  - P(WWDG)  \cdot P(EWDG) $$
$$ P(EWDG \cup WWDG) =  0.9775  $$
Let $P(EWDG  \cup  WWDG \cup SWDG) = $ probability of external hardware watchdog or windowed watchdog or software watchdog successfully recovering from a fault
\begin{align}
P(EWDG  \cup  WWDG \cup SWDG) =  P(EWDG \cup WWDG) + P(SWDG) \notag\\ - P(EWDG \cup WWDG) \cdot P(SWDG) \notag
\end{align}
$$  P(EWDG  \cup  WWDG \cup SWDG) =  0.9933 $$
Each task in FreeRTOS has a probability of at least $  P(EWDG  \cup  WWDG \cup SWDG) =  0.9933 $ for successfully recovering from a fault. The probability of a failure is very low as there are many layers of watchdogs. As there many SEUs per hour in LEO, layers of watchdogs ensure the OBC will survive throughout the mission duration.

\section*{Conclusion}
CubeSats are a low-cost alternative to full-sized satellites. As all CubeSats have a payload, an OBC is required to execute the payload. A single OBC has a MCU and some memory. Moreover, the OBC is built using COTS components as they are readily available and are cheap. However, COTS components are less tolerant to radiation, temperature, and vibrations when compared to space grade components. On the other hand, this paper presents hardware and software solutions to the problems. COTS components such as  MR4A16BCMA35, TMS570, and TPL5010-Q1 were chosen for the OBC. These COTS components have a higher tolerance to radiation, temperature, and vibrations. Moreover, these components are low-cost as they cost \$60, \$40, and \$2 respectively.

The MR4A16BCMA35 stores the program data and the sensor data. MR4A16BCMA35's TID of $4 \times 10^{4}$ rad (Si) is larger than the required TID of $2.825 \times 10^{3}$ rad (Si). Therefore, MR4A16BCMA35 survives the radiation in LEO.  OBC also uses TMS570, a MCU for CubeSat control. TMS570 has dual lockstep CPUs which protect against SEUs. Moreover, TMS570's TID of $5 \times 10^{3}$ rad (Si) meets the required TID of $2.825 \times 10^{3}$ rad (Si). Software solutions were also developed with the hardware solutions. A file system was designed to protect the data from radiation damage. In conclusion, these solutions can be used to increase the reliability of spacecraft and could be applied to radioactive environments.

\newpage
\bibliography{ArbutusReviewBibliography}
\bibliographystyle{apacite}

\end{document}